\documentclass[aps,prd,showkeys,showpacs,twocolumn,a4paper]{revtex4-1}
\usepackage{bm}
\usepackage{ulem}
\usepackage{color}
\usepackage{graphicx}
\usepackage{appendix}
\usepackage{epsfig}
\usepackage[colorlinks=true,linkcolor=blue]{hyperref}
\usepackage{epstopdf}
\usepackage{amsmath}
\usepackage{subfig}
\usepackage{amsmath}
\usepackage{comment}
\usepackage[dvipsnames]{xcolor}
\newcommand{\be}{\begin{equation}}
\newcommand{\ee}{\end{equation}}
\newcommand{\ba}{\begin{eqnarray}}
\newcommand{\ea}{\end{eqnarray}}

\begin{document}

\title{Thermoelectric behaviour of hot collisional and magnetized QCD medium from an effective kinetic theory}
\author{Manu Kurian}
\email{manu.kurian@iitgn.ac.in}
\affiliation{Indian Institute of Technology Gandhinagar, Gandhinagar-382355, Gujarat, India}


\begin{abstract}

The thermoelectric behaviour of quark-gluon plasma has been studied within the framework of an effective kinetic theory by adopting a quasiparticle model to incorporate the thermal medium effects. The thermoelectric response of the medium has been quantified in terms of the Seebeck coefficient. The dependence of the collisional aspects of the QCD medium on the Seebeck coefficient has been estimated by utilizing relaxation time approximation and Bhatnagar-Gross-Krook collision kernels in the effective Boltzmann equation. The thermoelectric coefficient is seen to depend on the quark chemical potential and collision aspects of the medium. Besides, the thermoelectric effect has been explored in a magnetized medium and the respective transport coefficients, such as magnetic field-dependent Seebeck coefficient and Nernst coefficient, have been estimated. The impacts of hot QCD medium interactions incorporated through the effective model and the magnetic field on the thermoelectric responses of the medium have been observed to be more prominent in the temperature regimes not very far from the transition temperature.

\end{abstract}



\maketitle

 \section{Introduction}
 Experimental programs at Relativistic Heavy Ion Collider (RHIC) and Large
Hadron Collider (LHC) have confirmed the existence of the hot and dense nuclear matter known as the quark-gluon plasma (QGP)~\cite{STAR,Aamodt:2010pb}. The space-time evolution of the QGP has been successfully described within the framework of relativistic dissipative hydrodynamics~\cite{Gale:2013da,Jaiswal:2016hex}. The non-equilibrium physics of the created medium critically depends on its transport coefficients associated with the momentum, thermal and electric charge transport processes. Besides quantifying the dissipative processes and system responses to electromagnetic fields, these transport coefficients act as input parameters for the hydrodynamical approach. The relevance of transport parameters for the quantitative description of the measured observable has been explored in the collision experiments at the RHIC and LHC~\cite{Teaney:2003kp, Romatschke:2007mq,ALICE:2016kpq,Adam:2016izf}.

Recently, the LHC and RHIC have reported that the directed flow $v_1$ of $D$ and $\bar{D}^0$ mesons are about three orders of magnitude higher than that of the charged hadrons~\cite{Acharya:2019ijj,Adam:2019wnk}. These observations indicate the existence of a strong magnetic field in the initial stages of the heavy-ion collision. The theoretical estimations of the strength of the magnetic field in the primary stage of collision are in the order of $(1-15)m_\pi^2$~\cite{Skokov:2009qp,Deng:PRC2012} and have generated a wide enthusiasm. The impact of the magnetic field on anomalous transport phenomena~\cite{Fukushima:2008xe,Sadofyev:2010pr}, heavy quark dynamics~\cite{Fukushima:2015wck,Singh:2020fsj}, quarkonia suppression~\cite{Hasan:EPJC2017,Singh:PRD972018}, electromagnetic probes~\cite{Bandyopadhyay:PRD2016}, jet quenching~\cite{Li:2016bbh}, transport and thermodynamic of QCD medium~\cite{Bandyopadhyay,Koothottil:2018akg,Dey:2019vkn} has gained much attention recently. Studies have shown that the lifetime of the magnetic field in the QGP may depend on the medium properties~\cite{Tuchin:PRC882013,McLerran:NPA9292014}, which indicates that the magnetic field may persist in the medium for a longer time than expected. Owing to the fact that the decay of the magnetic field is not completely modeled yet, the hot QCD medium properties have been explored both in the strong and weak magnetic field regimes. In a strongly magnetized medium $\sqrt{q_fB}\gg T$, where $q_fB$ is the strength of the field and $T$ is the temperature of the medium, the charged fermion motion is constrained along the direction of the magnetic field via Landau quantization. On the other hand, in the presence of a weak magnetic field $\sqrt{q_fB}\ll T$, cyclotron frequency captures the magnetic field effects.
 
The temperature gradient over the separation of fluid results in the thermal dissipation in the hot QCD medium. The thermal transport in the QGP medium has been explored within kinetic theory framework~\cite{Mitra:2017sjo,Kalikotay:2019fle}, effective models~\cite{Marty:2013ita,Kadam:2017iaz,Deb:2016myz,Greif:2013bb} and Kubo formalism~\cite{FernandezFraile:2009mi}. Similarly, the electromagnetic responses of the QGP can be quantified in terms of the electrical conductivity of the medium. Several works have been devoted to understand the electrical and thermal transport processes in a magnetized QCD medium~\cite{Feng:PRD962017,Das:2019ppb,Das:2019pqd,Thakur:2019bnf,Ghosh:2019ubc,Chatterjee:2019nld,Rath:2019vvi,K:2020yub}. Notably, the magnetic field induces anisotropy in the transport processes and in the strong field limit, the dominant contribution is from the longitudinal component of the transport coefficients associated with the thermal and electric charge transport processes. Thermoelectric or Seebeck effect describes the phenomena in which the temperature gradient generates the electric current in an electrically conducting medium and vice versa. The thermoelectric behaviour has been well studied for condensed matter systems~\cite{G,H}. In a recent study~\cite{F}, a related coefficient, namely, the Seebeck coefficient, has been investigated for a dense hadronic medium. The thermoelectric behaviour of the quark matter has started receiving attention very recently~\cite{Abhishek:2020wjm}. It is important to note that a finite quark chemical potential is required to obtain a non-zero thermoelectric current in the medium. Further, the presence of the magnetic field leads to anisotropic thermoelectric transport process in the medium. The thermoelectric effect in the presence of the magnetic field can be studied in terms of magnetic field-dependent Seebeck coefficient and Nernst coefficient in the weakly magnetized QGP medium. Transport coefficients characterizing the thermoelectric behaviour of the magnetized QCD medium, Seebeck and Nernst coefficients, has been estimated in both weak field~\cite{Das:2020beh,Zhang:2020efz} and strong field regimes~\cite{Dey:2020sbm}. It would be an interesting aspect to study the thermoelectric responses of the interacting QCD medium while including the effects of collisions and realistic equation of state (EoS) for both the cases of vanishing and non-vanishing magnetic fields. This sets the motivation for our investigations.

In the current analysis, the thermoelectric responses of the interacting 2-flavor QGP medium and the associated transport coefficients have been estimated within an effective quasiparticle model. The gluonic, quarks, and antiquarks degrees of freedom are modeled by utilizing the effective fugacity quasiparticle model (EQPM)~\cite{Chandra:2011en,Kurian:2017yxj}. The knowledge of system away from the equilibrium can be obtained by solving the consistently developed effective Boltzmann equation within the EQPM~\cite{Mitra:2018akk}. The collisional aspects of the interacting medium are incorporated in the analysis through relaxation time approximation (RTA) and Bhatnagar-Gross-Krook (BGK) collision kernels. The RTA turns out to be a viable approach to explore the dissipative process and the associated transport coefficients of the QCD medium~\cite{Czajka:2017wdo,Plumari:2012ep,Jaiswal:2014isa,Panda:2020zhr}. In a recent study~\cite{Bhadury:2020ngq}, the authors have derived dissipative hydrodynamics employing the RTA while including the transitions between quarks and gluons. The BGK kernel is the improvement over the RTA such that it conserves the particle number instantaneously. The formalism of the thermoelectric responses has been extended to a magnetized nuclear matter in terms of the magnetic field-dependent Seebeck coefficient and Hall-type Nernst coefficient. In a recent work~\cite{Kurian:2020qjr}, we have studied the relative significance of electric charge transport in the presence of an external electric field and thermal transport in a weakly magnetized medium in terms of the Wiedemann-Franz law. The present analysis is to understand the physics of generated electric field due to the temperature gradient in the collisional hot QCD medium. The dependence of magnetic field and collisions on the temperature behaviour of thermoelectric coefficients has been studied while embedding the realistic EoS effects.

The manuscript is organized as follows. The mathematical formulation of thermoelectric transport in a collisional hot QCD medium within the EQPM effective kinetic theory is presented in section II. Section III describes the thermoelectric behaviour and the associated transport coefficients of a magnetized QGP. Section IV is devoted to the results and the followed discussions. Finally, in section V, the present analysis is summarized with an outlook.

{\bf Notations and conventions:} The fractional charge of the up and down quarks are $q_f=2e/3, -e/3$, respectively. We define $u^{\mu}$ as the normalized unit vector with $u^{\mu}u_{\mu}=1$ and $\Delta^{\mu\nu}\equiv g^{\mu\nu}-u^\mu u^\nu$ as the projection operator orthogonal to $u^{\mu}$ where $g^{\mu\nu}=$diag$(1 ,-1, -1, -1)$ is the  metric tensor. The index $k$ represents the particle species. In the absence of magnetic field, the quantity $g_k$ denotes the degeneracy factor with $g_{q,\bar{q}}=\sum_f2N_c$ where $\sum_f$ is the sum over flavor $f$. We consider $N_f=2$ and massless limit of quarks in the present analysis.

\section{Effective description of thermoelectric effect in thermal QCD medium}
The diffusion of the charge carriers due to the temperature gradient in a medium result in the generation of the electric field, and the thermoelectric behaviour of the medium can be quantified with the associated transport parameter, the Seebeck coefficient. The effective description of the thermoelectric behaviour of the collisional QCD medium requires knowledge of the non-equilibrium part of the distribution function along with the proper modeling of the system at equilibrium. To that end, we employ the EQPM description of the medium. The particle four flow $N^{\mu}$ can be described within the EQPM as follows~\cite{Mitra:2018akk},
\begin{align}\label{1}
&N^{\mu}=g_q\int{\frac{d^3\mid{\bf{\Tilde{p}}}_q\mid}{(2\pi)^3\omega_{q}}\,\Tilde{p}_q^{\mu}\,\Big(f_q(x,\Tilde{p}_{q})-f_{\bar{q}}(x,\Tilde{p}_{\bar{q}})}\Big)\nonumber\\
&+\delta\omega_q\,g_q\int{\frac{d^3\mid{\bf{\Tilde{p}}}_q\mid}{(2\pi)^3\omega_{q}}\,\frac{\langle\Tilde{p}_q^{\mu}\rangle}{\epsilon_q}\, \Big(f_q(x,\Tilde{p}_{q})-f_{\bar{q}}(x,\Tilde{p}_{\bar{q}})}\Big),
\end{align}
where $\langle\Tilde{p}^{\mu}_q\rangle\equiv\Delta^\mu_\nu\,\Tilde{p}^{\nu}_q$ and ${{\Tilde{p}^{\mu}}}_q=(\omega_q, \bf{\Tilde{p}}_q)$ is the dressed four-momentum of the quasiquark in the medium. For the system not very far from local equilibrium, the EQPM momentum distribution function can be defined as,
\begin{align}\label{2}
&f_k=f^0_k+\delta f_k, &&f^0_{k}=\frac{z_k \exp{[-\beta (u\!\cdot\! p_k-a_k \mu)]}}{1 \pm z_k\exp{[-\beta (u\!\cdot\! p_k -a_k \mu)]}},
\end{align}
with ${\mid\delta f_k\mid} \ll {f^0_k}$, $\mu$ as the quark chemical potential, and $a_k=1, -1, 0$ for quarks, antiquarks and gluons, respectively. Here, $f^0_k$ is the EQPM equilibrium distribution function and $\delta f_k$ non-equilibrium part of the momentum distribution. The parameter $z_k$ is the temperature-dependent effective fugacity that encodes the medium interactions through lattice EoS. The quasiparticle four-momenta are related to the bare particle momenta $p_k^{\mu}=(\epsilon_k, {\bf p}_k)$ through $z_k$. The effective fugacity parameter modifies the energy dispersion as follows,
\begin{equation}\label{3}
\Tilde{p_k}^{\mu} = p_k^{\mu}+\delta\omega_k\, u^{\mu}, \qquad
\delta\omega_k= T^{2}\,\partial_{T} \ln(z_{k}),
\end{equation}
such that the quasiparticle energy $\omega_k=\epsilon_k+\delta\omega_k$ where $\epsilon_k=\mid{\bf{\Tilde{p}}}_k\mid$. It is important to note that the temperature behaviour of quark and antiquark effective fugacities remains the same as $z_k$ are not associated with any conserved current in the medium, $i.e.$, $z_q=z_{\bar{q}}$. The net baryon density $n$ can be defined from the zeroth component of the $N^{\mu}$ and takes the following form in the massless limit,
\begin{align}\label{4}
n =\frac{T^3}{\pi^2}\sum_{k} g_ka_k\mathrm{PolyLog}~[3,-z_qe^{-\alpha_k}],
\end{align}
where $\alpha_k=\beta a_k\mu$. Similarly, the microscopic definition of current density in the QCD medium within the EQPM has the form as follows,
\begin{align}\label{5}
{ j}^i=\sum_kg_k\int{d\Tilde{P}_k\,q_{f_k}{ v}^i_k\,\delta f_k}-\sum_k\delta\omega_kg_k\int{d\Tilde{P}_k\,q_{f_k}\frac{{ v}^i_k}{\epsilon_k}\,\delta f_k},
\end{align}
with $d\Tilde{P}_k=\frac{d^3\mid{\bf{\Tilde{p}}}_k\mid}{(2\pi)^3}$ and $v^i=\frac{\Tilde{p}^i}{\Tilde{p}^0}$. The system away from equilibrium $\delta f_k$ can be described by the relativistic Boltzmann equation. The effective Boltzmann within the EQPM takes the following form,
\begin{equation}\label{6}
\Tilde{p}^{\mu}_k\,\partial_{\mu}f_k(x,\Tilde{p}_k)+\bigg(F_k^{\mu}\left(u\!\cdot\!\tilde{p}_k\right)+q_{f_k}F^{\mu\nu}\Tilde{p}_{k\,  \nu}\bigg)\partial^{(\Tilde{p}_k)}_{\mu} f_k 
= C[f_{k}],
\end{equation}
where $C[f_{k}]$ is the collision kernel that quantifies the rate of change of distribution functions due to the collisional processes in the medium. Here, $F^{\mu\nu}$ represents the  electromagnetic  field  strength  tensor and $F_k^{\mu}=-\partial_{\nu}(\delta\omega_k u^{\nu}u^{\mu})$ denotes the mean field force term that arises from the basic conservation laws in the medium. We solve the Boltzmann equation Eq.~(\ref{6}) to obtain $\delta f_k$ with a proper choice of the collision kernel.

\subsection{Seebeck coefficient within RTA}

Within the RTA, the collisional aspects in the QCD medium can be described in terms of thermal relaxation time $\tau_{R_k}$ as follows,
\begin{align}\label{A.1}
  &C_k =-(u.\Tilde{p}_k)\frac{\delta f_k}{\tau_{R_k}}.
\end{align}
The relaxation time for the $2\rightarrow 2$ elastic scattering processes in the medium has been included in the present analysis~\cite{Anderson_Witting}. Notably, in a strongly magnetized medium $1\rightarrow 2$ processes such as quark-antiquark annihilation and vice versa are kinematically possible and considered as the dominant processes in the medium~\cite{Hattori:2017qih}. In the massless case, the relaxation time for quarks/antiquarks takes the following form~\cite{Hosoya:1983xm,Thakur:2019bnf},
\begin{align}\label{7}
    \tau_{R} =\frac{1}{5.1T\alpha_{\text{eff}}^2 \ln (\frac{1}{\alpha_{\text{eff}}})\Big(1+0.12(2N_f +1)\Big)},
\end{align}
where $\alpha_{\text{eff}}$ is the effective coupling that can be realized as the charge renormalization within the EQPM description of hot QCD medium and is related to running coupling constant  $\alpha_{s}(T, \mu)$ as follows,
\begin{align}\label{8}
\frac{\alpha_{\text{eff}}}{\alpha_{s}(T, \mu)}=&
\Bigg[ \dfrac{2N_c}{\pi^{2}}
\mathrm{PolyLog}\,[2,z_{g}]-\dfrac{2N_f}
{\pi^{2}}\mathrm{PolyLog}\,[2,-z_{q}]
\nonumber\\
&+\mu^2 \frac{N_f}{\pi^2}\frac{z_q}{1+z_q}\Bigg] \dfrac{1}{\Big( \frac{N_c}{3}+\frac{N_f}{6}+\mu^2\frac{N_f}{2\pi^2}\Big)}.
\end{align}
Substituting Eq.~(\ref{A.1}) in Eq.~(\ref{6}) and employing an iterative Chapman-Enskog like approach~\cite{Jaiswal:2013npa}, we solve the Boltzmann equation while considering the force term as ${\bf F}=q_{f_k}{\bf E}$ in the case of a finite electric field. Hence, the first order correction to the EQPM momentum distribution can be expressed as,
\begin{align}\label{9}
\!\!\delta f_k =& \tau_{R_k}\Bigg[\frac{1}{T} \bigg\{\Tilde{p}_k^0~\partial_0 T + \Tilde{p}_k^i\,\partial_i T \bigg\} \!+\! \frac{T}{\Tilde{p}^0_k} \!\bigg\{\Tilde{p}^0_k\,\partial_0\alpha_k \nonumber\\
&\!+\! \Tilde{p}_k^i\,\partial_i \alpha_k \bigg\}\!-\! \frac{1}{\Tilde{p}^0_{k}} \!\bigg\{\Tilde{p}^0_{k}\, \Tilde{p}_{k}^\nu\,\partial_0 u_\nu \!+\! \Tilde{p}^i_{k}\, \Tilde{p}_{k}^\nu\,\partial_i u_\nu \bigg\}\! \nonumber\\
&+\! \theta\,\delta\omega_k-q_{f_k}\bar{\bf v}_k.{\bf E}\Bigg]\frac{\partial f^0_k}{\partial\epsilon_k},
\end{align}
where $\bar{\bf v}_k={\bf v}_k\frac{\omega_k}{\epsilon_k}$. Here, $\theta=\partial_\mu u^\mu$ and the traceless part of the velocity gradient $\langle\langle\partial_\mu u_\nu \rangle\rangle$ act as the source terms for the bulk and shear viscous force in the medium. Simplifying Eq.~(\ref{9}) by Gibbs-Duhem relation  $\partial_i \Big(\frac{\mu}{T}\Big)=-\frac{h}{T^2}(\partial_iT-\frac{T}{nh}\partial_iP)$ where $h=\frac{\varepsilon+P}{n}$, $\varepsilon$ and $P$ are the enthalpy, energy density and pressure of the system respectively, we obtain,
\begin{align}\label{10}
\!\!\delta f_k = \tau_{R_k}\frac{\partial f^0_k}{\partial\epsilon_k}\bigg[\big(\omega_{k}-a_kh\big){ v}^i_k\frac{{{\partial_i}} T}{T} -q_{f_k}\bar{ v}^i_k{ E^i}\bigg].
\end{align}
It is important to emphasize that in the steady state the momentum conservation gives $\partial_i P=0$. The Eq.~(\ref{10}) denotes the non-equilibrium correction to the distribution function due to the temperature gradient and electric field in the medium. Substituting Eq.~(\ref{10}) in the electric current density as described in Eq.~(\ref{5}), we obtain
\begin{align}\label{11}
{\bf j}&=\sum_k\frac{g_k\tau_{R_k}q_{f_k}}{3T}\int{d\Tilde{P}_k\,\frac{\mid{\bf{\Tilde{p}}}_k\mid^2}{\omega_k^2}}\Big(\omega_{k}-a_kh\Big)\frac{\partial f^0_k}{\partial\epsilon_k}{\bm{\nabla} T}\nonumber\\
&-\sum_k\delta\omega_k\frac{g_k\tau_{R_k}q_{f_k}}{3T}\int{d\Tilde{P}_k\,\frac{\mid{\bf{\Tilde{p}}}_k\mid}{\omega_k^2}}\Big(\omega_{k}-a_kh\Big)\frac{\partial f^0_k}{\partial\epsilon_k}\bm{\nabla} T\nonumber\\
&-\sum_k\frac{g_k\tau_{R_k}q^2_{f_k}}{3}\int{d\Tilde{P}_k\,\frac{\mid{\bf{\Tilde{p}}}_k\mid^2}{\omega_k}}\frac{1}{\epsilon_k}\frac{\partial f^0_k}{\partial\epsilon_k}{\bf E}\nonumber\\
& +\sum_k\frac{g_k\tau_{R_k}q^2_{f_k}}{3}\int{d\Tilde{P}_k\,\frac{\mid{\bf{\Tilde{p}}}_k\mid^2}{\omega_k}}\frac{1}{\epsilon_k}\frac{\partial f^0_k}{\partial\epsilon_k}{\bf E}.
\end{align}
The thermoelectric effect can be described by setting up ${\bf j}=0$ in a steady state, and hence we have,
\begin{equation}\label{12}
  {\bf E}=S{\bm{\nabla}}T, 
\end{equation}
such that the generated electric field is proportional to the temperature gradient which can be quantified in terms of the Seebeck coefficient $S$. Employing Eq.~(\ref{11}) and Eq.~(\ref{12}), and performing the thermodynamic integrals within the EQPM description, the Seebeck coefficient takes the form as follows,
\begin{equation}\label{13}
S=\frac{I_1(T, \mu)}{I_2(T, \mu)},
\end{equation}
where $I_1$ and $I_2$ can be defined in terms of $PolyLog$ functions as,
\begin{align}
I_1&=-\frac{1}{{6\pi^2}}\sum_k{g_k\tau_{R_k}q_{f_k}}\Bigg[-6T^2\mathrm{PolyLog}\,[3,-e^{\alpha_k}z_{k}]\nonumber\\
&+4\delta\omega_kT\mathrm{PolyLog}\,[2,-e^{\alpha_k}z_{k}]+{a_khT}\nonumber\\
&\Big(2\mathrm{PolyLog}\,[2,-e^{\alpha_k}z_{k}]+3\frac{\delta\omega_k}{T}\mathrm{Log}\,[1+e^{\alpha_k}z_{k}]\Big)\Bigg],\\
I_2&=\frac{1}{3\pi^2}\sum_k{g_k\tau_{R_k}q^2_{f_k}}\Bigg[T^2\mathrm{PolyLog}\,[2,-e^{\alpha_k}z_{k}]\nonumber\\
&+\delta\omega_kT\mathrm{Log}\,[1+e^{\alpha_k}z_{k}]\Bigg].
\end{align}
It is important to note that the term $I_2$ is related to the electrical conductivity of the QGP medium, $i.e.$, $I_2=-\sigma_{e}$ in the case of a vanishing magnetic field. The temperature behaviour of the $\sigma_{e}$ and the Seebeck coefficient critically depends on the thermal relaxation time. The effect of collisions can be further studied with the BGK collisional term and could be thought of as an improvement over the RTA results.

\subsection{Seebeck coefficient within BGK kernel}

The BGK collisional aspects are observed to have a significant impact on the collective modes, refractive index, and electric charge transport process of the QCD medium~\cite{Schenke:2006xu,Kumar:2017bja,Khan:2020rdw,Jiang:2016dkf,Carrington:2003je}. The BGK collision term takes the form as follows,
\begin{equation}\label{15}
C[f_k]=-\nu \Big[f_k-\frac{N_k}{N_k^0}f^0_k\Big]\\
 =-\nu \Big[\delta f_k -\frac{f^0_k}{N_k^0} \int d\Tilde{P}_k\, \delta f_k\Big],
\end{equation}
where $\nu$ is the collisional frequency which is independent of the momentum of particles and
\begin{align}\label{16}
&N_k= \int d\Tilde{P}_k\,  f_k, && N_k^0 =\int d\Tilde{P}_k\,f^0_k,
\end{align}
denote the particle densities of the $k-$th species. The BGK collision kernel preserves number conservation instantaneously, unlike conventional RTA integral $i.e.$, $\int d\Tilde{P}_k\, C[f_k] =0.$ Note that in the limit $\frac{N_k}{N_k^0}=1$, the BGK collisional term  reduces to RTA kernel. Substituting Eq.~(\ref{15}) in Eq.~(\ref{6}) and solving the Boltzmann equation, we have
\begin{align}\label{17}
\delta f_{k}=&\delta {f_{k}}^{(0)}+i\nu(iD)^{-1}\dfrac{f^0_k}{N_k^0}\Bigg[\int{d\Tilde{P}^{'}_k\,}
{\delta f_{k}^{(0)}(\Tilde{p}^{'},X)}\Bigg]\nonumber\\
&+i\nu(iD)^{-1}\dfrac{f^0_k}{N_k^0}\dfrac{i\nu}{N_k^0}\Bigg[\int{d\Tilde{P}^{'}_k\,}{(iD)^{-1} f^0_k}\nonumber\\&\times\int{d\Tilde{P}^{''}_k\,}
{\delta f_{k}^{(0)}(\Tilde{p}^{''},X)}\Bigg]\, +...... ,
\end{align}
where $D=v.\partial_X+\nu$ with $v=(1, {\bf v}_k)$ and the RTA equivalent form of the part of distribution function away from equilibrium $\delta f_{k}^{(0)}$ takes the following form,
\begin{align}\label{18}
\!\!\delta f^{(0)}_k = \nu^{-1}\,\frac{\Tilde{p}_k^i}{\Tilde{p}_k^0}\bigg[\big(\omega_{k}-a_kh\big)\frac{\partial_iT}{T}-\frac{\Tilde{p}_k^0}{\epsilon_k}q_{f_k}E^i\bigg]\frac{\partial f^0_k}{\partial\epsilon_k}.
\end{align}
Employing Eq.~(\ref{17}) in Eq.~(\ref{5}) and keeping terms in leading order of $\nu^{-1}$, the current density takes the following form,
\begin{widetext}
\begin{align}\label{19}
{ j}^i=&\sum_k{g_k\nu^{-1}q_{f_k}}\int{d\Tilde{P}_k\,v_k^i}\Bigg\{\bigg[\frac{\Tilde{p}^i_k}{\omega_{k}}\Big(\omega_{k}-a_kh\Big)\frac{\partial_iT }{T}-\frac{\Tilde{p}^i_k}{\epsilon_k}q_{f_k}E^i\bigg]\frac{\partial f^0_k}{\partial\epsilon_k}+\frac{f^0_k}{N^0_k}\nu^{-1}
\int{d\Tilde{P}^{'}_k\,}\nonumber\\&\times\bigg[\frac{\Tilde{p}^{'\,i}_k}{\omega^{'}_{k}}\Big(\omega^{'}_{k}-a_kh\Big)\frac{\partial_iT }{T}-\frac{\Tilde{p}^{'\,i}_k}{\epsilon^{'}_k}q_{f_k}E^i\bigg]\frac{\partial f^0_k}{\partial\epsilon^{'}_k}\Bigg\}+\sum_k{\delta\omega_kg_k\nu^{-1}q_{f_k}}\int{d\Tilde{P}_k\,\frac{v_k^i}{\mid{\bf{\Tilde{p}}}_k\mid}}\Bigg\{\bigg[\frac{\Tilde{p}^i_k}{\omega_{k}}\Big(\omega_{k}-a_kh\Big)\frac{\partial_iT }{T}\nonumber\\&-\frac{\Tilde{p}^i_k}{\epsilon_k}q_{f_k}E^i\bigg]\frac{\partial f^0_k}{\partial\epsilon_k}+\frac{f^0_k}{N^0_k}\nu^{-1}
\int{d\Tilde{P}^{'}_k\,}\bigg[\frac{\Tilde{p}^{'\,i}_k}{\omega^{'}_{k}}\Big(\omega^{'}_{k}-a_kh\Big)\frac{\partial_iT }{T}-\frac{\Tilde{p}^{'\,i}_k}{\epsilon^{'}_k}q_{f_k}E^i\bigg]\frac{\partial f^0_k}{\partial\epsilon^{'}_k}\Bigg\}.
\end{align}
\end{widetext}
Setting up ${\bf j}=0$ by employing Eq.~(\ref{19}) and performing the thermodynamic integrals, the Seebeck coefficient within the BGK collision kernel can be defined as,
\begin{equation}\label{20}
S=\frac{K_1(T, \mu)}{K_2(T, \mu)},
\end{equation}
\begin{widetext}
\begin{align}\label{21}
K_1=&-\frac{1}{6\pi^2}\sum_k{\frac{g_kq_{f_k}}{\nu}}\Bigg[-6T^2\,\mathrm{PolyLog}\,[3,-e^{\alpha_k}z_{k}]+4\delta\omega_kT\,\mathrm{PolyLog}\,[2,-e^{\alpha_k}z_{k}]+{a_khT}\Big(2\,\mathrm{PolyLog}\,[2,-e^{\alpha_k}z_{k}]\nonumber\\&+3\frac{\delta\omega_k}{T}\mathrm{Log}\,[1+e^{\alpha_k}z_{k}]\Big)-2\frac{A_k}{N_k^0}\Big(-T^2\,\mathrm{PolyLog}\,[3,-e^{\alpha_k}z_{k}]+\delta\omega_kT\,\mathrm{PolyLog}\,[2,-e^{\alpha_k}z_{k}]\Big)\Bigg],\\
K_2=&\frac{1}{3\pi^2}\sum_k{\frac{g_kq^2_{f_k}}{\nu}}\Bigg[\Big(T^2\,\mathrm{PolyLog}\,[2,-e^{\alpha_k}z_{k}]+\delta\omega_kT\,\mathrm{Log}\,[1+e^{\alpha_k}z_{k}]\Big)+\frac{B_k}{N_k^0}\Big(-T^2\,\mathrm{PolyLog}\,[3,-e^{\alpha_k}z_{k}]\nonumber\\&+\delta\omega_kT\,\mathrm{PolyLog}\,[2,-e^{\alpha_k}z_{k}]\Big)\Bigg],\label{21.1}
\end{align}
\end{widetext}
where $A_k$ and $B_k$ respectively take the forms as follows,
\begin{align}\label{22}
    A_k&=\frac{1}{2\pi^2}\Bigg[6T^3\,\mathrm{PolyLog}\,[3,-e^{\alpha_k}z_{k}]-{a_khT^2}\nonumber\\&\times\Big(2\,\mathrm{PolyLog}\,[2,-e^{\alpha_k}z_{k}]+\frac{\delta\omega_k}{T}\mathrm{Log}\,[1+e^{\alpha_k}z_{k}]\Big)\Bigg],\\
    B_k&=\frac{T^3}{\pi^2}\,\mathrm{PolyLog}\,[2,-e^{\alpha_k}z_{k}].
\end{align}
It is important to note that in the limit of $z_k=1$, the analysis reduces back to that in the case of ideal EoS. The term $\mid K_2\mid$ denotes the electrical conductivity of the QGP within the BGK collision kernel. Let us now proceed to discuss the thermoelectric effect in a magnetized QCD medium.

\section{Thermoelectric responses of a magnetized QGP within the EQPM}

In a weakly magnetized medium, the temperature is the dominant energy scale in comparison with the strength of the magnetic field. Hence, the energy dispersion of quarks/antiquarks remains intact in the presence of the weak magnetic field, unlike the Landau level dispersion of charged fermions in a strongly magnetized medium. In the presence of a weak magnetic field, the Boltzmann equation has the form,
\begin{align}\label{23}
\Big[-\frac{\partial f^0_k}{\partial\epsilon_k}&\big({\omega_{k}-a_kh}\big)\frac{1}{T}({\bf v}_k.\bm{\nabla} T)\Big] \nonumber\\ &+ q_{f_k} \Big[{\bf E}+({\bf v}_k\times {\bf B})\Big].\frac{\partial f_k}{\partial{\bf{\Tilde{p}}}_k}=-\frac{\delta f_k}{\tau_{R_k}}.
\end{align}
We consider that the dependence of the magnetic field on the thermal relaxation time for the binary scattering process is entering through the one-loop magnetic field-dependent coupling constant $\alpha_s(q_fB, T)$. The form of $\alpha_s(q_fB, T)$ for a weakly magnetized medium is presented in Refs.~\cite{Bandyopadhyay,Ayala:2018wux}. However, in the presence of a strong magnetic field, the relaxation time critically depends on the strength of the magnetic field as the $1\rightarrow 2$ processes are kinematically possible in the medium~\cite{Hattori:2017qih}. The non-equilibrium part of the quasiparton distribution in the presence of a weak magnetic can be obtained by solving the Boltzmann equation. To that end, we choose the following ansatz,
\begin{align}\label{24}
\delta f_k=({\bf{\Tilde{p}}}_k\, .\, {\bf \Xi})\,\frac{\partial f^0_k}{\partial\epsilon_k},
\end{align}
where ${\bf \Xi}$ is defined as,
\begin{align}\label{25}
{\bf \Xi}=&\alpha_1\, {\bf E}+\alpha_2\, {\bf b}+\alpha_3\, \big({\bf E}\times{\bf b}\big)+\alpha_4\, {\bm{\nabla} }T\nonumber\\&+\alpha_5\, \big({\bm{\nabla} }T\times{\bf b}\big)+\alpha_6\, \big({\bm{\nabla} }T\times{\bf E}\big),
\end{align}
with unit vector ${\bf b}=\frac{{\bf B}}{\mid{\bf B}\mid}$ represents the direction of the magnetic field. In general, there will be other independent force terms other than the thermal driving force and Lorentz force in the Boltzmann equation that corresponds to the momentum transport and the associated viscous coefficients in the medium. Different components of the transport parameters associated with the electric charge, thermal and momentum transport processes in the magnetized medium are studied in Ref.~\cite{Dash:2020vxk}. Since the Lorenz force vanishes in the equilibrium case as $\frac{\partial f^0_k}{\partial{\bf{\Tilde{p}}}_k}\propto {\bf v}_k$, the Eq.~(\ref{23}) reduces to the following form,
\begin{align}\label{26}
&\bigg[-\big({\omega_{k}-a_kh}\big)\frac{1}{T}({\bf v}_k.\bm{\nabla} T)+q_{f_k}(\bar{\bf v}_k\, .{\bf E})\bigg] \frac{\partial f^0_k}{\partial\epsilon_k}\nonumber\\ &+q_{f_k} \bigg[({\bf v}_k\times {\bf B}).\, {\bf \Xi}\bigg]\frac{\partial f^0_k}{\partial\epsilon_k}=-\frac{1}{\tau_{R_k}}({\bf{\Tilde{p}}}_k\, .\, {\bf \Xi})\,\frac{\partial f^0_k}{\partial\epsilon_k}.
\end{align}
The unknown parameters $\alpha_{i}, (i=1, 2,.., 6)$  in the non-equilibrium part of the distribution can be obtained by substituting Eq.~(\ref{24}) in Eq.~(\ref{26}). Hence, we obtain
\begin{widetext}
\begin{align}\label{27}
&-\big({\omega_{k}-a_kh}\big)\frac{1}{T}({\bf v}_k.\bm{\nabla} T)+q_{f_k}(\bar{\bf v}_k\, .{\bf E})+q_{f_k}\alpha_1 ({\bf v}\times {\bf B})\,.{\bf E}+q_{f_k}\alpha_2 ({\bf v}\times {\bf B})\,.{\bf b}+ q_{f_k}\alpha_3 ({\bf v}\times {\bf B})\,.({\bf E}\times {\bf b})\nonumber\\ &+q_{f_k}\alpha_4 ({\bf v}\times {\bf B})\,.{\bm{\nabla} }T+q_{f_k}\alpha_5 ({\bf v}\times {\bf B})\,.({\bm{\nabla}}T\times {\bf b})+q_{f_k}\alpha_6 ({\bf v}\times {\bf B})\,.({\bm{\nabla}}T\times {\bf E}) \nonumber\\ &=-\frac{\omega_k}{\tau_{R_k}}\bigg[\alpha_1{\bf v}_k.{\bf E}+\alpha_2{\bf v}_k.{\bf b}+\alpha_3{\bf v}_k.({\bf E}\times{\bf b})+\alpha_4{\bf v}_k.{\bm{\nabla}} T+\alpha_5{\bf v}_k.({\bm{\nabla}}T\times{\bf b})+\alpha_6{\bf v}_k.({\bm{\nabla}}T\times{\bf E})\bigg].
\end{align}
\end{widetext}
Comparing different independent tensor structures on both sides of Eq.~(\ref{27}), we obtain the following relations for the parameters $\alpha_{i}, (i=1, 2,.., 6)$, 
\begin{align}\label{28}
 -\frac{\omega_k}{\tau_{R_k}}\alpha_1&=\frac{\omega_k}{\epsilon_k}q_{f_k}+\alpha_3q_{f_k} \mid{\bf B}\mid,\\
- \frac{\omega_k}{\tau_{R_k}}\alpha_2&=-\alpha_3q_{f_k} ({\bf B}.{\bf E})-\alpha_5q_{f_k} ({\bf B}.{\bm{\nabla}}T),\\
 -\frac{\omega_k}{\tau_{R_k}}\alpha_3&=-\alpha_1q_{f_k} \mid{\bf B}\mid,\\
- \frac{\omega_k}{\tau_{R_k}}\alpha_4&=-\frac{\big(\omega_k-a_kh\big)}{T}+\alpha_5q_{f_k} \mid{\bf B}\mid,\\
 -\frac{\omega_k}{\tau_{R_k}}\alpha_5&=-\alpha_4q_{f_k} \mid{\bf B}\mid,\label{28.1}
\end{align}
with $\alpha_6=0$. For the quantitative analysis of the thermoelectric effect in a magnetized medium, we consider the magnetic field direction along the $z-$axis and the directions of the electric and temperature gradient in the $x-y$ plane in the present analysis, $i.e.$, $({\bf b}.{\bf E})=({\bf b}.{\bm{\nabla}}T)=0$. Further solving the Eqs.~(\ref{28})-(\ref{28.1}), the parameters take the forms as follows,
\begin{align}\label{29}
\alpha_1&=-\frac{\tau_R}{\epsilon_k}\frac{q_{f_k}}{(1+\tau_{R_k}^2\, \Omega^2_{c\, k})},\\
\alpha_2&=0,\\
\alpha_3&=-\frac{\tau_{R_k}^2}{\epsilon_k}\frac{q_{f_k}}{(1+\tau_{R_k}^2\, \Omega^2_{c\, k})}\,  \Omega_{c\, k},\\
\alpha_4&=\frac{\tau_{R_k}}{\omega_k}\frac{1}{T}\frac{\big({\omega_{k}-a_kh}\big)}{(1+\tau_{R_k}^2\, \Omega^2_{c\, k})},\\
\alpha_5&=\frac{\tau_{R_k}^2}{\omega_k}\frac{1}{T}\frac{\big({\omega_{k}-a_kh}\big)}{(1+\tau_{R_k}^2\, \Omega^2_{c\, k})}\,  \Omega_{c\, k},\label{29.1}
\end{align}
where $\Omega_{c\, k}=\frac{q_{f_k}\mid B\mid}{\omega_k}$ denotes the cyclotron frequency in the presence of the magnetic field. Employing Eqs.~(\ref{29})-(\ref{29.1}) in Eq.~(\ref{24}) we obtain,
\begin{align}\label{30}
&\delta f_k=\frac{\tau_{R_k}}{(1+\tau_{R_k}^2\, \Omega^2_{c\, k})}\Bigg[-q_{f_k}\Big(\bar{\bf v}_k.{\bf E}+\tau_{R_k}\, \Omega_{c\, k} {\bf v}_k. \big({\bf E}\times {\bf b}\big)\Big)\nonumber\\
&+\big({\omega_{k}-h_k}\big)\frac{1}{T}\Big(\big({\bf v}_k.{\bm{\nabla}}T\big)+\tau_{R_k}\, \Omega_{c\, k} {\bf v}_k. \big({\bm{\nabla}}T\times {\bf b}\big)\Big)\Bigg]\,\frac{\partial f^0_k}{\partial\epsilon_k}.
\end{align}
By substituting Eq.~(\ref{30}) in Eq.~(\ref{5}), the component of the current density along the $x-$axis can be obtained as follows,
\begin{align}\label{31}
&j_{x}=\sum_{k}\frac{g_k\tau_{R_k}q_{f_k}}{3}\int d\Tilde{P}_k\,\frac{v_k^2}{(1+\tau_{R_k}^2\, \Omega^2_{c\, k})}\bigg[-\frac{\omega_k}{\epsilon_k}q_{f_k}\Big( E_x\nonumber\\
&+\tau_{R_k}\, \Omega_{c\, k}E_y\Big)+\big({\omega_{k}-a_kh}\big)\frac{1}{T}\Big(\frac{dT}{dx}+\tau_{R_k}\, \Omega_{c\, k} \frac{dT}{dy}\Big)\bigg]\,\frac{\partial f^0_k}{\partial\epsilon_k}\nonumber\\&
-\sum_{k}\delta\omega_k\frac{g_k\tau_{R_k}q_{f_k}}{3}\int d\Tilde{P}_k\,\frac{v_k^2}{\epsilon_k(1+\tau_{R_k}^2\, \Omega^2_{c\, k})}\bigg[-q_{f_k}\Big( E_x\nonumber\\
&+\tau_{R_k}\, \Omega_{c\, k}E_y\Big)+\big({\omega_{k}-a_kh}\big)\frac{1}{T}\Big(\frac{dT}{dx}+\tau_{R_k}\, \Omega_{c\, k} \frac{dT}{dy}\Big)\bigg]\,\frac{\partial f^0_k}{\partial\epsilon_k}.
\end{align}
Similarly, the component of current density along the $y-$axis take the following form,
\begin{align}\label{32}
&j_{y}=\sum_{k}\frac{g_k\tau_{R_k}q_{f_k}}{3}\int d\Tilde{P}_k\,\frac{v_k^2}{(1+\tau_{R_k}^2\, \Omega^2_{c\, k})}\bigg[-\frac{\omega_k}{\epsilon_k}q_{f_k}\Big( E_y\nonumber\\
&-\tau_{R_k}\, \Omega_{c\, k}E_x\Big)+\big({\omega_{k}-a_kh}\big)\frac{1}{T}\Big(\frac{dT}{dy}-\tau_{R_k}\, \Omega_{c\, k} \frac{dT}{dx}\Big)\bigg]\,\frac{\partial f^0_k}{\partial\epsilon_k}\nonumber\\&
-\sum_{k}\delta\omega_k\frac{g_k\tau_{R_k}q_{f_k}}{3}\int d\Tilde{P}_k\,\frac{v_k^2}{\epsilon_k(1+\tau_{R_k}^2\, \Omega^2_{c\, k})}\bigg[-q_{f_k}\Big( E_y\nonumber\\
&-\tau_{R_k}\, \Omega_{c\, k}E_x\Big)+\big({\omega_{k}-a_kh}\big)\frac{1}{T}\Big(\frac{dT}{dy}-\tau_{R_k}\, \Omega_{c\, k} \frac{dT}{dx}\Big)\bigg]\,\frac{\partial f^0_k}{\partial\epsilon_k}.
\end{align}
The charge carriers travel along the temperature gradient ($x-y$ plane) generate an electric field that induces an electric current in the opposite direction, and in the steady state the net current vanishes in the medium. By setting up $j_x=0$ and $j_y=0$, we can represent the generated electric field in terms of the temperature gradient. Hence, from Eq.~(\ref{31}) and  Eq.~(\ref{32}) we have,
\begin{align}\label{33}
    &E_x=S_B\, \frac{dT}{dx}+N\mid{\bf B}\mid\, \frac{dT}{dy},\\
    &E_y=S_B\, \frac{dT}{dy}-N\mid{\bf B}\mid\, \frac{dT}{dx},\label{33.1}
\end{align}
where $S_B$ and $N$ are the Seebeck and Nernst coefficients associated with the thermoelectric transport process in the presence of the magnetic field. In the case of vanishing magnetic field, the Eq.~(\ref{33}) and Eq.~(\ref{33.1}) reduces to Eq.~(\ref{12}) such that the coefficient $S_B$ reduces to $S$ and Nernst coefficient vanishes in the medium. The $S_B$ and $N$ take the following forms,
\begin{align}\label{34}
    &S_B= \frac{L_1L_3+L_2L_4}{L_1^2+L_2^2},
    &&N\mid{\bf B}\mid=\frac{L_1L_4-L_2L_3}{L_1^2+L_2^2},
\end{align}
where $L_{i}, i=(1, 2, 3, 4)$ are the thermodynamic integrals in the presence of the magnetic field and can be defined as follows,
\begin{widetext}
\begin{align}\label{35}
L_1=&\frac{1}{3}\sum_{k}g_k\tau_{R_k}q^2_{f_k}\Bigg\{\int{d\Tilde{P}_k}\frac{\mid{\bf{\Tilde{p}}}_k\mid^2}{\omega_k}\frac{1}{\epsilon_k}\frac{1}{(1+\tau_{R_k}^2\, \Omega^2_{c\, k})}\frac{\partial f^0_k}{\partial\epsilon_k}-\delta\omega_k\int{d\Tilde{P}_k}\frac{\mid{\bf{\Tilde{p}}}_k\mid^2}{\omega_k}\frac{1}{\epsilon^2_k}\frac{1}{(1+\tau_{R_k}^2\, \Omega^2_{c\, k})}\frac{\partial f^0_k}{\partial\epsilon_k}\Bigg\},\\
L_2=&\frac{1}{3}\sum_{k}g_k\tau_{R_k}q^2_{f_k}\Bigg\{\int{d\Tilde{P}_k}\frac{\mid{\bf{\Tilde{p}}}_k\mid^2}{\omega_k}\frac{1}{\epsilon_k}\frac{\tau_{R_k}\, \Omega_{c\, k}}{(1+\tau_{R_k}^2\, \Omega^2_{c\, k})}\frac{\partial f^0_k}{\partial\epsilon_k}-\delta\omega_k \int{d\Tilde{P}_k}\frac{\mid{\bf{\Tilde{p}}}_k\mid^2}{\omega_k}\frac{1}{\epsilon^2_k}\frac{\tau_{R_k}\, \Omega_{c\, k}}{(1+\tau_{R_k}^2\, \Omega^2_{c\, k})}\frac{\partial f^0_k}{\partial\epsilon_k}\Bigg\},\\
L_3=&\frac{1}{3T}\sum_{k}g_k\tau_{R_k}q_{f_k}\Bigg\{\int{d\Tilde{P}_k}\frac{\mid{\bf{\Tilde{p}}}_k\mid^2}{\omega^2_k}\frac{\big({\omega_{k}-a_kh}\big)}{(1+\tau_{R_k}^2\, \Omega^2_{c\, k})}\frac{\partial f^0_k}{\partial\epsilon_k}-\delta\omega_k\int{d\Tilde{P}_k}\frac{\mid{\bf{\Tilde{p}}}_k\mid}{\omega^2_k}\frac{\big({\omega_{k}-a_kh}\big)}{(1+\tau_{R_k}^2\, \Omega^2_{c\, k})}\frac{\partial f^0_k}{\partial\epsilon_k}\Bigg\},\\
L_4=&\frac{1}{3T}\sum_{k}g_k\tau_{R_k}q_{f_k}\Bigg\{\int{d\Tilde{P}_k}\frac{\mid{\bf{\Tilde{p}}}_k\mid^2}{\omega^2_k}\frac{\tau_{R_k}\, \Omega_{c\, k}\big({\omega_{k}-a_kh}\big)}{(1+\tau_{R_k}^2\, \Omega^2_{c\, k})}\frac{\partial f^0_k}{\partial\epsilon_k}-\delta\omega_k\int{d\Tilde{P}_k}\frac{\mid{\bf{\Tilde{p}}}_k\mid}{\omega^2_k}\frac{\tau_{R_k}\, \Omega_{c\, k}\big({\omega_{k}-a_kh}\big)}{(1+\tau_{R_k}^2\, \Omega^2_{c\, k})}\frac{\partial f^0_k}{\partial\epsilon_k}\Bigg\}.
\end{align}
\end{widetext}
The integrals $L_2$ and $L_4$ vanishes at ${\bf B}=0$, and hence Nernst coefficient vanishes in the absence of a magnetic field. 

In a strongly magnetized medium, the motion of fermions is constrained in the direction of the magnetic field. Various transport processes in the QCD medium have been initially studied in the lowest Landau level (LLL) approximation ($T^{2}\ll \mid q_{f_k} B\mid$)~\cite{Hattori:2017qih} and later on the more realistic regime $gT\ll \sqrt{\mid q_fB\mid}$ where higher Landau level contributions are significant~\cite{Kurian:2018qwb}. The longitudinal current density $j_z$ due to the $1+1-$dimensional Landau level dynamics of the charged particles in the presence of the strong magnetic field has been studied in Ref.~\cite{Fukushima:2017lvb}. Following the same formalism and invoking $j_z=0$, the Seebeck coefficient in the strongly magnetized medium $S_B^{\parallel}$ can be defined as,
\begin{align}\label{36}
  &{\bf E}=S_B^{\parallel}{\bm{\nabla}}T, &&S^{\parallel}_B=\frac{I^{\parallel}_1(q_fB, T, \mu)}{I^{\parallel}_2(q_fB, T, \mu)},
\end{align}
where the integrals takes the form,
\begin{align}\label{37}
I^{\parallel}_1=& -\sum_{l=0}^{\infty}\sum_{k}g_l\dfrac{\mid q_{f_k}B\mid}{2\pi}
\dfrac{N_c}{T^2}\int_{-\infty}^{\infty}{\dfrac{d\Tilde{p}_{z_k}}
{2\pi}\tau_{\text{eff}}\dfrac{(\omega^l_{k}-a_kh^l)}
{\omega^{l\, 2}_{k}}}\nonumber\\
&\times\Tilde{p}_{z_k}^2
f^{l\,0}_k(1-f^{l\,0}_k)\nonumber\\
&+\sum_{l=0}^{\infty}\sum_{k}g_l\delta\omega_k\dfrac{\mid {q_{f_k}}B\mid}
{2\pi}\dfrac{N_c}{T^2}\int_{-\infty}^{\infty}{\dfrac{d\Tilde{p}_{z_k}}
{2\pi}\tau_{\text{eff}}\dfrac{(\omega^l_{k}-a_kh^l)}
{\omega^{l\, 2}_{k}}}\nonumber\\
&\times\dfrac{\Tilde{p}_{z_k}^2}{E_l}
f^{l\,0}_k(1-f^{l\,0}_k).
\end{align}
Here, $g_l=(2-\delta_{l0})$ is the spin degeneracy with $l$ as the Landau level, $h^l$ is the enthalpy in the strongly magnetized QGP, and $\omega^l_{k}=E_l+\delta\omega_k$ with $E_l=\sqrt{p_{z}^{2}+2l\mid q_{f_k}B\mid}$. The EQPM particle distribution function in a strong magnetic field takes the form as,
\begin{equation}\label{38}
f^{l\,0}_{k}=\dfrac{z_{k}\exp{\Big[-\beta \Big(E_l-a_k\mu\Big)\Big]}}{1+ z_{k}\exp{\Big[-\beta \Big(E_l-a_k\mu\Big)\Big]}}.
\end{equation}
Similarly, the integral $I^{\parallel}_2$ can be defined as,
\begin{align}\label{39}
    I^{\parallel}_2=-\sigma_{\parallel}, 
\end{align}
where $\sigma_{\parallel}$ is the longitudinal electrical conductivity within the EQPM as described in Ref.~\cite{Kurian:2019fty}. Here, $\tau_{\text{eff}}$ is the thermal relaxation time for dominant $1\rightarrow 2$ processes in the strongly magnetized medium~\cite{Hattori:2017qih}.

\begin{figure*}
 \centering
 \subfloat{\includegraphics[scale=0.4]{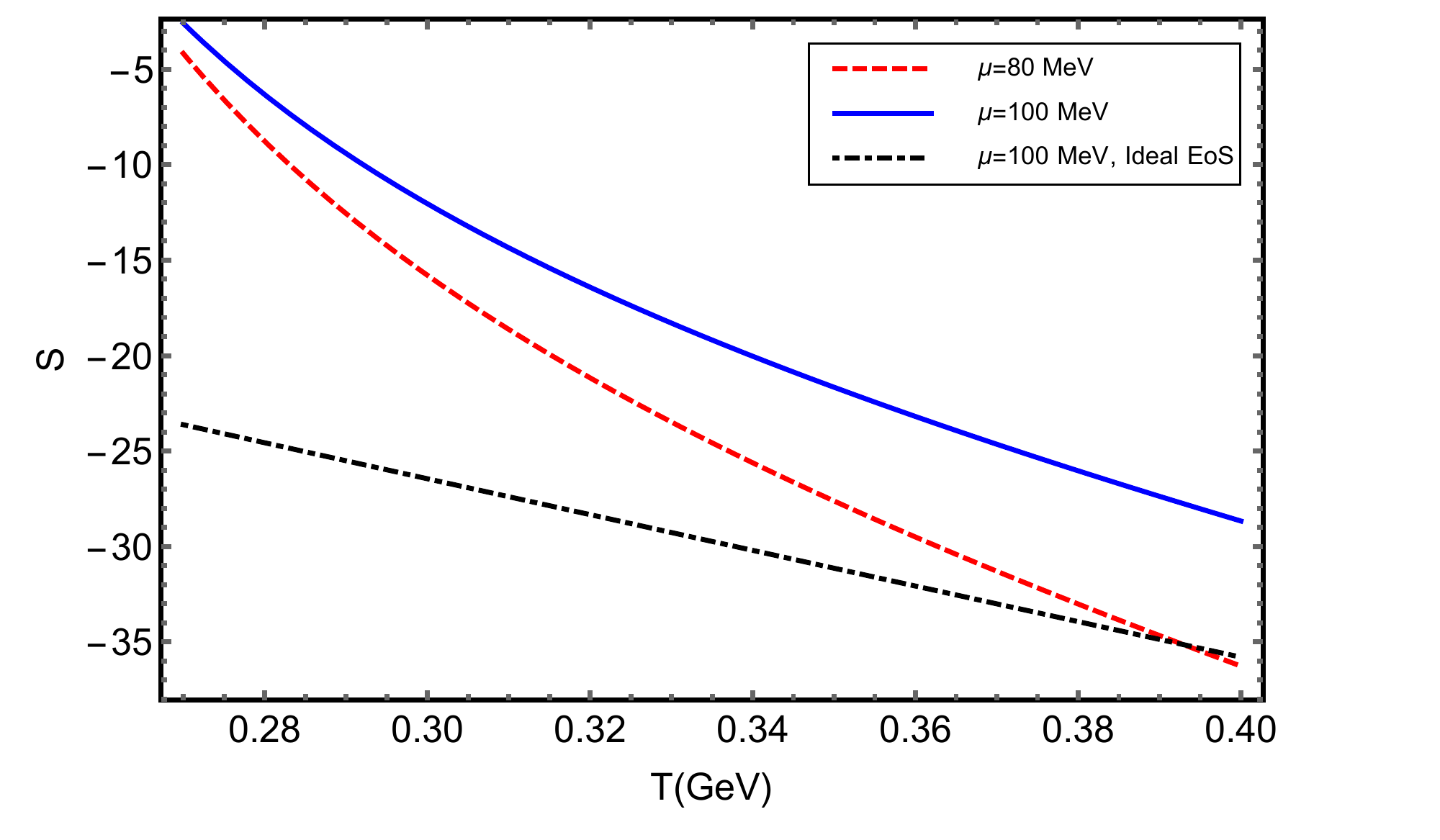}}
 \subfloat{\includegraphics[scale=0.4]{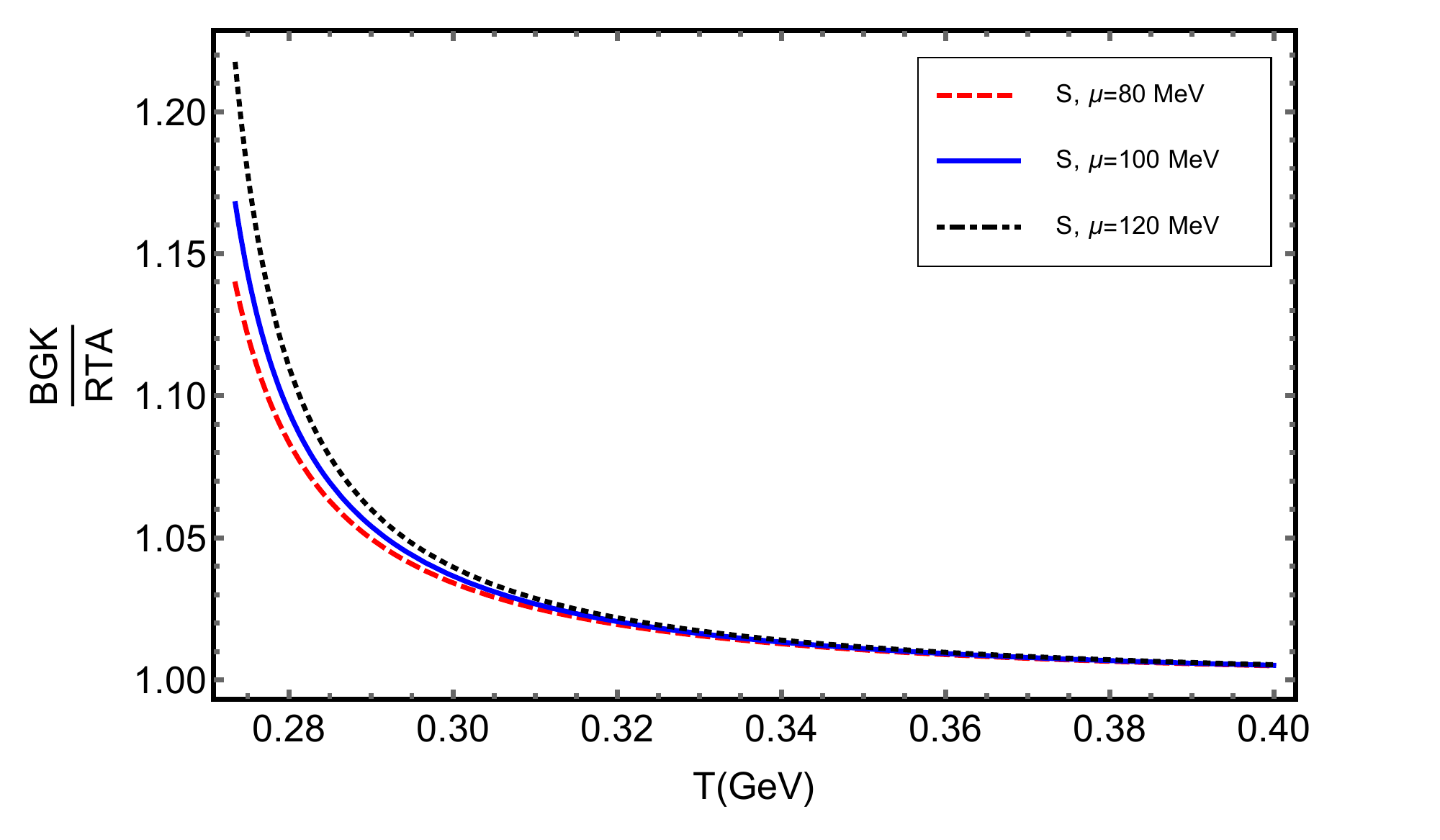}}
\caption{(Left panel) Dependence of chemical potential and medium interactions on the temperature behaviour of Seebeck coefficient of the 2-flavor QGP within the RTA. (Right panel) The RTA results are compared with the estimations with the BGK collision kernel. }
\label{f1}
\end{figure*}


\section{Results and discussions}

We initiate the discussions with the temperature dependence of the Seebeck coefficient in the 2-flavor QGP medium. The dependence of quark chemical potential and thermal medium interactions on the temperature behaviour of the thermoelectric coefficient is depicted in Fig.~\ref{f1} (left panel) in the case of vanishing magnetic field within the RTA. It is observed that the mean field contributions that arise from the hot medium interactions to the Seebeck coefficient are more pronounced in the temperature regimes not very far from the transition temperature $T_c$. In a previous study~\cite{Kurian:2020qjr}, we have studied the relative behaviour of electric and thermal transport in the medium in terms of the Wiedemann-Franz law and observed that the law is violated with the inclusion of EoS effects in temperature regimes near $T_c$. In a similar line, we have observed that the hot QCD medium interactions significantly modify the low temperature behaviour of the thermoelectric effect in the medium. The negative sign of $S$ is arising from the term $(\omega_k-a_kh)$ for the 2-flavor system within the EQPM, in contrast with the term $(\omega_k-a_kh)^2$ while describing the Lorentz number $L=\frac{\kappa}{\sigma_eT}$ where $\kappa$ is the thermal conductivity of the medium. At very high temperature, the EQPM results reduce back to the results of non-interacting medium owing to the fact that $z_{k}\rightarrow 1$ and $\delta\omega_k\rightarrow 0$ at the asymptotic limit. At the massless and ideal EoS limits, the coefficient $S$ vanishes for the 3-flavor QGP at ${\bf B}=0$. Along with the temperature gradient in the medium, a finite quark chemical potential is also required for the thermoelectric effect in the hot QCD medium~\cite{Abhishek:2020wjm}. This is attributed to the fact that in the QGP medium, there are positive and negative charge carriers for the transport process, unlike in the condensed matter system. The quark chemical potential is seen to have a strong dependence on the thermoelectric effect in the medium. The coefficient decreases with an increase in the quark chemical potential at a particular temperature. This observation is in line with the results of hot and dense hadron gas as described in Ref.~\cite{F}.

The collisional aspects of the QGP medium are embedded in the analysis through the RTA and BGK kernels. In Fig.~\ref{f1} (right panel), the ratio of BGK description of the Seebeck coefficient to the RTA result is plotted as a function of temperature. The BGK description of the Seebeck coefficient is described in Eq.~(\ref{20}). The first term of Eq.~(\ref{21}) and Eq.~(\ref{21.1}) describe the RTA results, and other terms denote further corrections within the BGK description. For the quantitative estimation, the parameter $\nu$ is assumed as the thermal average of inverse of relaxation time for the elastic collisions. In the absence of a magnetic field, the choice of $\nu$ will not depend on the temperature behaviour of the thermoelectric coefficient. It is observed that the collisional effect to the thermoelectric coefficient is more prominent in the temperature regime near $T_c$. The current density has contributions from the temperature gradient (with non-zero quark chemical potential) in the medium and also from the generated electric field. Both the contributions get corrections within the BGK analysis in comparison with the RTA result as described in Eq.~(\ref{13}). Notably, in the massless and ultra-relativistic limit ($z_k\rightarrow 1$) limit, $S$ will be independent of the choice of collision integral, as the corrections to both contributions to the current density cancel exactly each other while defining the Seebeck coefficient.

\begin{figure}[h]
\hspace{-1cm}
  \subfloat{\includegraphics[scale=0.4]{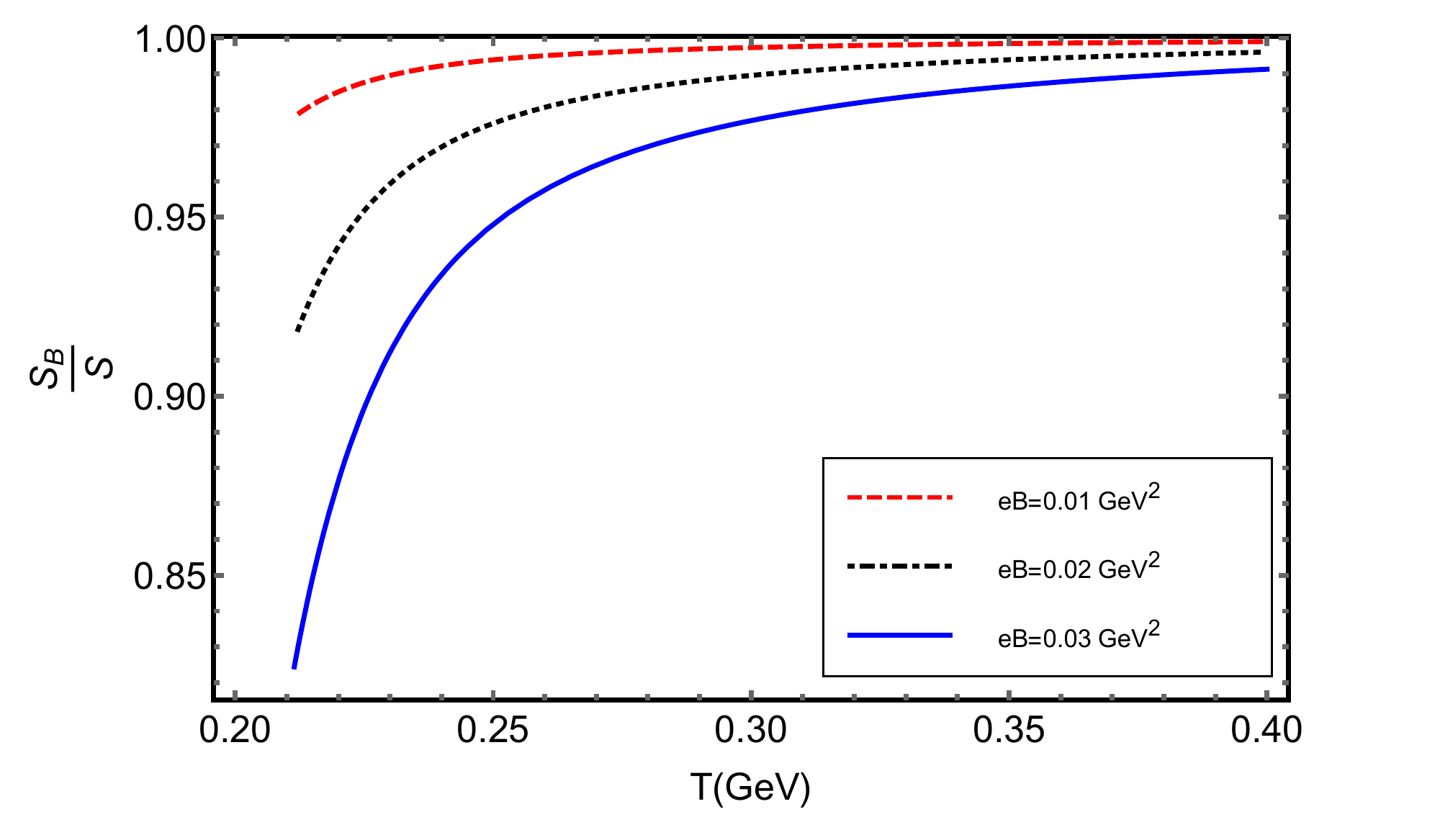}}
\caption{ Temperature dependence of Seebeck coefficient at $\mu=100$ MeV in a weakly magnetized medium.}
\label{f2}
\end{figure}
\begin{figure*}
 \centering
 \subfloat{\includegraphics[scale=0.4]{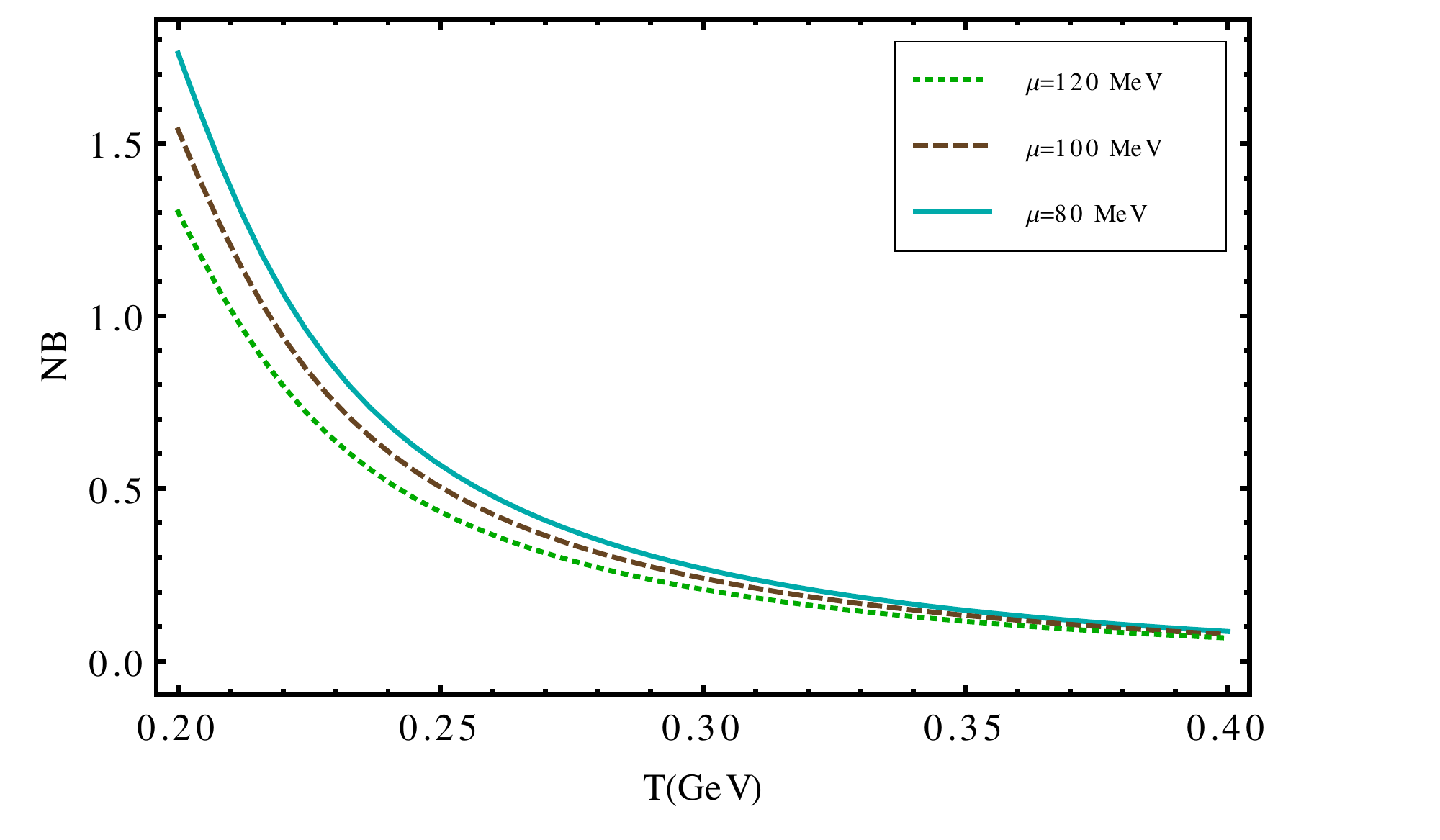}}
 \subfloat{\includegraphics[scale=0.4]{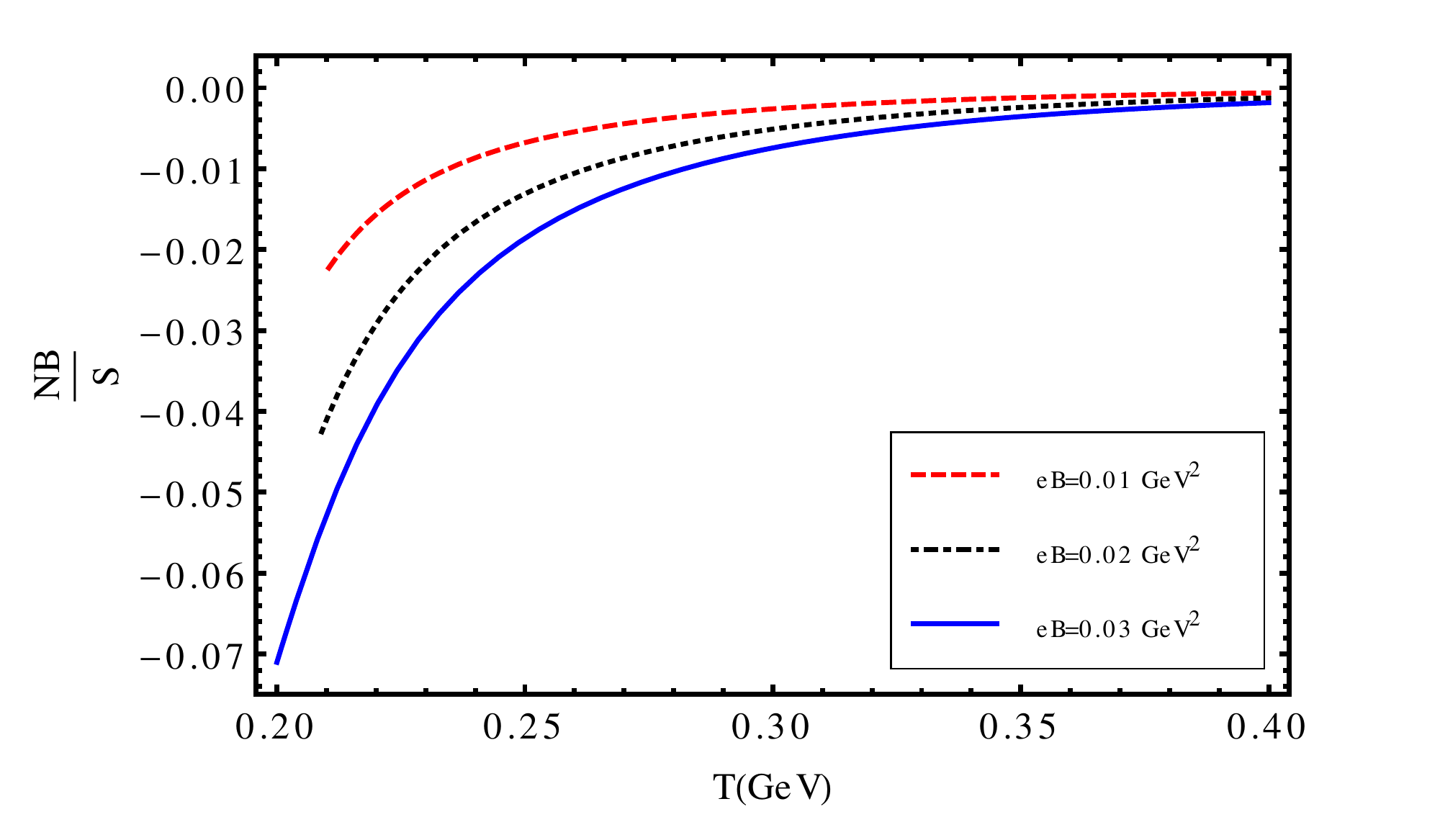}}
\caption{(Left panel) Dependence of chemical potential on the temperature behaviour of Nernst coefficient at $\mid eB\mid=0.03$ GeV$^2$. (Right panel) The impact of the magnetic field on Nernst coefficient at $\mu=100$ MeV.}
\label{f3}
\end{figure*}

The magnetic field induces anisotropy in the system and leads to magnetic field-dependent Seebeck coefficient and Nernst coefficient associated with the thermoelectric effect in the magnetized medium as described in Eq.~(\ref{33}) and Eq.~(\ref{33.1}). The impact of the magnetic field on the Seebeck coefficient is shown in Fig.~\ref{f2}. In the weakly magnetized medium, the magnetic field dependence of the thermoelectric coefficient is entering through the Lorentz force via cyclotron frequency $\mid \Omega_{c\, k}\mid$. The effect of the magnetic field is observed to be more significant in the lower temperature regimes. As the temperature increases, the particle motion is completely dominated by the temperature, and the effect of the magnetic field vanishes. However, in the temperature regimes near $T_c$ the Seebeck coefficient decreases with an increase in the strength of the field due to the factor $\frac{1}{1+\tau_{R_k}^2\, \Omega^2_{c\, k}}$. It is important to emphasize that the Eq.~(\ref{34}) reduces back to Eq.~(\ref{12}) at the limit of ${\bf B}=0$.

The motion of the charged fermion gets deflected in the presence of a weak magnetic field due to the Lorentz force. The Hall-type conductivity associated with the thermal and electric charge transport in the hot QCD has been explored in Refs.~\cite{Kurian:2020qjr,Das:2019pqd}. The temperature variation of the Hall-type transport coefficient associated with the thermoelectric effect, Nernst coefficient, in the magnetized medium is depicted in Fig.~\ref{f3}. In contrast to the Seebeck coefficient, the Nernst coefficient is a positive quantity and is critically dependent on the quark chemical potential and strength of the magnetic field. It is seen that the $N\mid{\bf B}\mid$ decreases with an increase in chemical potential at a finite magnetic field. The dependence of the strength of the magnetic field on the Nernst coefficient is studied by plotting the ratio $\frac{N\mid{\bf B}\mid}{S}$ with the temperature in Fig.~\ref{f3} (right panel). The ratio approaches zero asymptotically, which indicates that the impact of the magnetic field on the thermoelectric transport is negligible at sufficiently high temperature in the weakly magnetized QGP. However, the magnetic field effects on the Nernst coefficient are visible in the low temperature regimes near $T_c$. Further, it is important to emphasize that the Nernst coefficient vanishes in a strongly magnetized medium due to the $1+1-$dimensional constraint motion of the charged particle.

\section{Conclusion and Outlook}

In this article, we have presented an analysis on the thermoelectric transport process and the associated transport coefficients in a collisional and magnetized hot QCD medium. The realistic EoS effects are embedded in the analysis within the framework of the EQPM through the temperature-dependent fugacity parameters. The non-equilibrium part of the momentum distribution of effective degrees of freedom is obtained by solving the effective transport equation within the EQPM employing an iterative Chapman-Enskog like approach while choosing a proper collision kernel. The electric field generated due to temperature gradient in the hot QCD medium at finite quark chemical potential is quantified in terms of the Seebeck coefficient. The temperature behaviour of the Seebeck coefficient in the collisional 2-flavor QGP medium has been investigated. The effects of collision are incorporated in the analysis through the RTA and BGK collision kernels. The thermal medium interactions of the QCD medium is seen to have a significant impact on the thermoelectric behaviour of the medium. The effects of the collisions and quark chemical potential are more visible in the temperature regimes near the transition temperature.

Further, we have studied the thermoelectric behaviour of a weakly magnetized QCD matter. In the analysis, the temperature is considered to be the dominant energy scale in comparison with the strength of the magnetic field. The magnetic field induces anisotropy in the thermoelectric behaviour of the medium. We have estimated the transport coefficients characterizing the thermoelectric behaviour of the magnetized QCD medium, magnetic field-dependent Seebeck and Nernst coefficients, within the effective transport equation. We have demonstrated the dependence of the magnetic field and quark chemical potential on thermoelectric coefficients in the weakly magnetized medium. The analysis is further extended to the strong magnetic field regime while considering the Landau level kinematics of the charged particle. Notably, the Nernst coefficient vanishes in a strongly magnetized medium due to the longitudinal motion of the particles. 

The study of thermoelectric behaviour of the collisional QCD medium in the presence of space-time varying magnetic fields would be an interesting direction to work. The induced electric field from the decay of the magnetic field may put a constraint in the generated electric field from the temperature gradient at a finite chemical potential. A very recent study~\cite{Liu:2020dxg} proposed that temperature and chemical potential gradient in the heavy-ion collision experiment may induce spin Hall current. These aspects will be taken up for future investigations.

\section*{acknowledgments}
The author would like to acknowledge Vinod Chandra for the immense encouragement, helpful discussions, and critical reading of the manuscript. The author acknowledges Indian Institute of Technology Gandhinagar for Institute postdoctoral fellowship and further records a deep sense of gratitude to the people of India for their generous support for the research in fundamental sciences.



{}

\end{document}